\documentclass[micromachines,article,submit,moreauthors,10pt,a4paper,pdftex]{mdpi} 
\firstpage{1} 
\articlenumber{x}
\doinum{10.3390/------}
\pubvolume{xx}
\pubyear{2017}
\copyrightyear{2017}
\externaleditor{Academic Editor: name}
\history{Received: date; Accepted: date; Published: date}

\usepackage{graphicx}
\usepackage{dcolumn}
\usepackage{bm}
\usepackage{color}
\usepackage{units}

\Title{Advanced Fabrication of Single-crystal Diamond Membranes for Quantum Technologies}

\Author{Michel Challier $^{1,\ddagger}$, Selda Sonusen $^{1,\ddagger}$, Arne Barfuss $^{2}$, Dominik Rohner $^{2}$, Daniel Riedel $^{2}$, Johannes Koelbl $^{2}$, Marc Ganzhorn $^{2}$, Patrick Appel $^{2}$, Patrick Maletinsky $^{2}$ and Elke Neu $^{1,}$*}

\AuthorNames{Michel Challier, Selda Sonusen, Arne Barfuss, Dominik Rohner, Daniel Riedel, Johannes Koelbl, Marc Ganzhorn, Patrick Appel, Patrick Maletinsky and Elke Neu}

\address{%
$^{1}$ \quad Universit\"at des Saarlandes, Fachrichtung Physik, 66123 Saarbr\"ucken, Germany\\
$^{2}$ \quad Department of Physics, University of Basel, Klingelbergstrasse 82, Basel CH-4056, Switzerland}

\corres{Correspondence: elkeneu@physik.uni-saarland.de; Tel.: +49-681-302-2739}

\secondnote{These authors contributed equally to this work.}

\abstract{Many promising applications of single crystal diamond and its color centers as sensor platform and in photonics require free-standing membranes with a thickness ranging from several micrometers to the few \unit[100]{nm} range. In this work, we present an approach to conveniently fabricate such thin membranes with up to about one millimeter in size. We use commercially available diamond plates (thickness \unit[50]{$\mu$m}) in an inductively coupled reactive ion etching process which is based on argon, oxygen and SF$_6$. We thus avoid using toxic, corrosive feed gases and add an alternative to previously presented recipes involving chlorine-based etching steps. Our membranes are smooth (RMS roughness < \unit[1]{nm}) and show moderate thickness variation (central part: < \unit[1]{$\mu$m} over $\approx \,$\unit[200x200]{$\mu$m$^2$}). Due to an improved etch mask geometry, our membranes stay reliably attached to the diamond plate in our chlorine-based as well as SF$_6$-based processes. Our results thus open the route towards higher reliability in diamond device fabrication and up-scaling.}


\begin{document}

\section{Introduction \label{seclabel:introduction}}
Single-crystal diamond (SCD) represents an outstanding material platform for quantum technologies including fields like sensing and photonics: For the latter, its bandgap of \unit[5.5]{eV} creates a wide transparency window, whereas its refractive index of $\approx \,$2.4\,\cite{Zaitsev2001} for visible light enables efficient light confinement, e.g.\ in photonic crystal cavities\,\cite{Aharonovich2014a} or in one-dimensional waveguides that enable tailored light-matter-interaction even on the single photon level \cite{Bhaskar2017}. In sensing applications, using chemically-stable, bio-compatible, ultra-hard SCD ensures reliable operation in various extreme environments. For radiation sensors, SCD additionally offers high carrier mobility, radiation hardness and breakdown voltage\,\cite{Pomorski2013}. In addition to the outstanding material properties of diamond itself, individual, optically-active point defects in diamond, so-called color centers, supply additional functionality, e.g.\ as single spin quantum systems. A prominent example is the nitrogen vacancy (NV) color center which serves as an extremely versatile sensor for magnetic and electric fields as well as temperature and strain\,\cite{Rondin2014,Bernardi2017,Dolde2011,Neumann2013,Teissier2014}. In this context, diamond enables high precision sensing as its main carbon isotope $^{12}$C has no nuclear spin. Consequently, electronic spins of color centers can have long coherence times even at room temperature rendering them  very sensitive, fully-stable sensors for fields external to the host crystal. However, only ultrapure SCD with high crystal quality consistently preserves those advantageous properties of color centers. 

Many of these promising applications of SCD and its color centers require free-standing membranes with a thickness ranging from several micrometers to the few \unit[100]{nm} range: Radiation and pressure sensing\,\cite{Pomorski2013,Momenzadeh2016} as well as radioisotope batteries\,\cite{Delfaure2016} require thin membranes. Color centers incorporated into thin SCD membranes have been coupled to cavities (e.g., 2D photonic crystal cavities\,\cite{RiedrichMoeller2015}, micro-Fabry-P\'erot cavities\,\cite{Riedel2017}) or to nano-mechanical structures\,\cite{Teissier2014}.  Thin membranes also form the starting point to fabricate SCD scanning probes\,\cite{Appel2016,Kleinlein2016, Maletinsky2012} that enable, e.g.\ magnetic sensing with nanoscale resolution\,\cite{Thiel2016}. A significant challenge for SCD applications arises from the fact that thin ($<$ \unit[50]{$\mu$m},\,\cite{Riedrichmoeller2011}) SCD films with good crystal quality cannot be grown on a non-diamond substrate. Only such growth would allow for straightforward creation of thin, free-standing membranes supported by a frame of substrate material by selectively (wet-)etching of the substrate. Consequently, convenient processes forming thin membranes from homoepitaxially-grown, high-quality SCD plates are of high technological relevance.  

In this work, we optimize a process to create few $\mu$m thin, smooth SCD membranes with up to about one millimeter lateral size starting from commercially available SCD plates (initial thickness of $\approx \,$\unit[50]{$\mu$m}). To achieve the thinning, we use inductively coupled reactive ion etching (ICP-RIE). Our process avoids etching the whole surface of the plate but rather aims for membrane windows (up to \unit[1x1]{mm$^2$}). Thus, the surrounding SCD plate is conserved as a holding frame and eases handling for subsequent fabrication of free standing structures like scanning probes, cantilevers or transferable micro-membranes. We report a simplified RIE process based on cycling steps using the non-toxic, non-corrosive gases argon, oxygen and SF$_6$. Together with an optimized shadow mask for RIE, our process reliably enables fabricating smooth SCD membranes in standard ICP-RIE machines. We compare different RIE recipes and characterize the membranes using scanning electron microscopy (SEM), atomic force microscopy (AFM) as well as laser scanning microscopy (LSM).

Previous work employed a lift-off process, where a $\mu$m-deep, buried layer in SCD is graphitized after ion bombardment\,\cite{Piracha2016a, Lee2013}. However, due to residual ion damage it is necessary to overgrow the lifted-off membrane with pristine SCD in a chemical vapor deposition (CVD) process. Subsequently, RIE removes the original material. Thus, RIE as well as customized SCD deposition are involved adding to the complexity of the process. Similar approaches based on focused ion beam milling of thin membranes result in low-quality, highly-contaminated membranes\,\cite{Babinec2011}. A very recent approach to free-standing devices is undercutting structures fabricated in single crystal diamond by using an isotropic plasma etch\,\cite{Wan2018}. First results with this approach indicate challenges concerning thickness gradients in the devices. We also point out that this approach cannot be used to fabricate membranes fully clamped on all sides.  
Previous work on fabricating thin membranes via RIE reported separation of the membranes from the surrounding thick plate due to inhomogeneous etching \,\,\cite{Appel2016}. Moreover, often chlorine(Cl$_2$)-based plasma chemistry was used\,\cite{Appel2016, Maletinsky2012,Tao2014,Teissier2014}. Though being highly-advantageous for ultra-smooth SCD etching\,\cite{Lee2008}, this chemistry requires avoiding plasma contact of silicon or SiO$_2$ parts as this would contaminate the etching process, leading to roughening of the membranes\,\cite{Appel2016}. SiO$_2$ and silicon parts are, however, commonly used in many ICP-RIE etching systems. Moreover, Cl$_2$-based plasma processes pose high restrictions concerning gas safety and handling, treatment of RIE exhaust gas as well as the mandatory use of a load-lock in the RIE reactor. 
      
\section{Shadow mask manufacturing and starting material \label{seclabel:mask_fab}}
\begin{figure}
	\centering
	\includegraphics[width=0.8\linewidth]{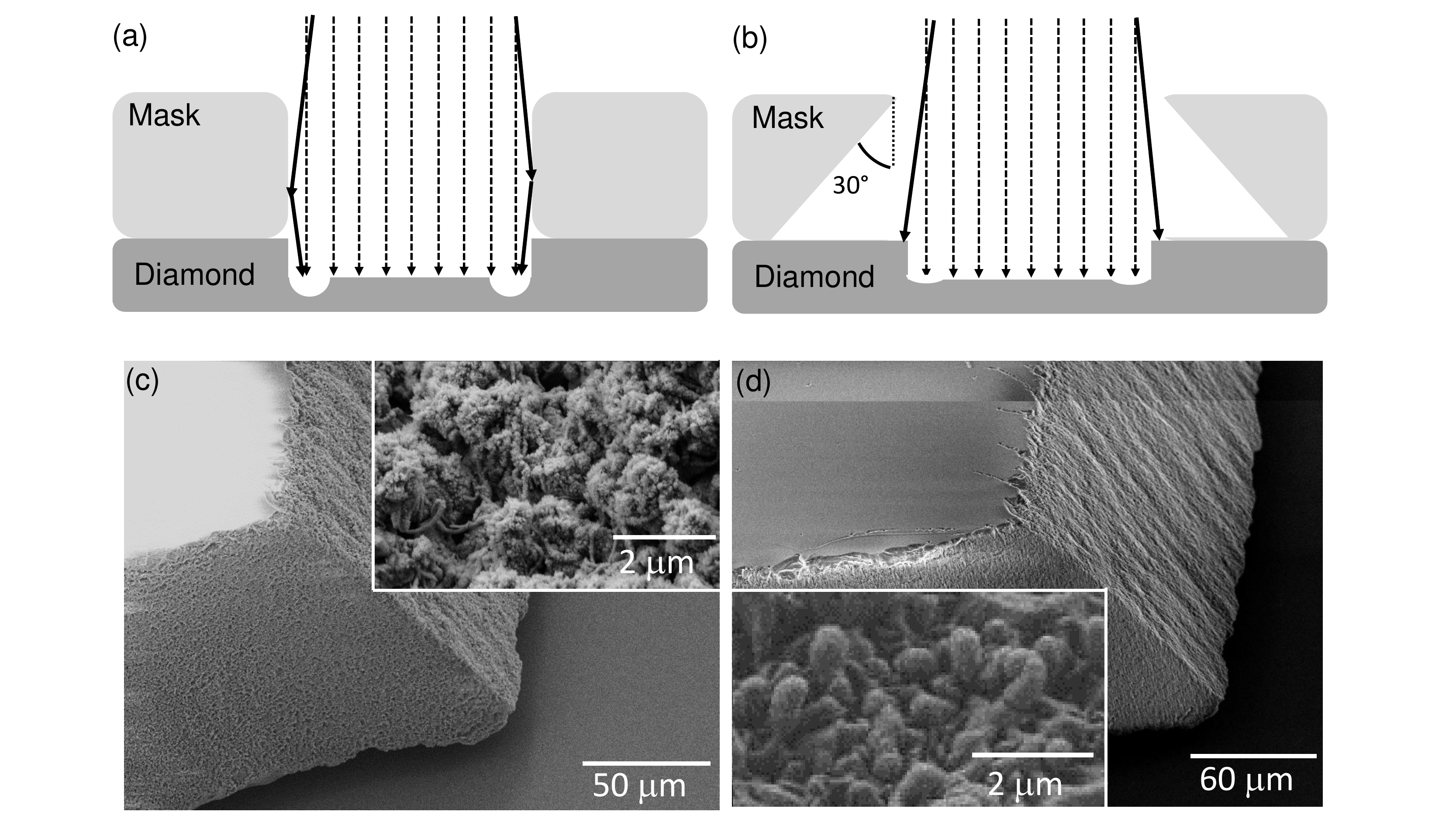}
	\caption{Quartz mask layout and morphology: (a) Formation of a trench close to the sidewalls: deflected ions (solid arrows) locally enhance the etch rate in comparison to regions that are only hit by non-deflected ions (dashed arrows). Part (b) depicts the improved mask layout and its effect on the trench formation. By cutting the mask's sidewalls under an angle of roughly 30$^{\circ}$, we avoid ion deflection on the mask's sidewalls. Laser-cut quartz masks before (c) and after RIE cleaning (d) of the laser cut edge [low magnification images: etching time \unit[30]{mins}, plasma parameters see Tab.\ \ref{tab:recipes}, SF$_6$/Ar cleaning (O)]. A change of morphology indicates the removal of the deposit in the laser cut as discernible from the insets [inset: etching time \unit[60]{mins}, plasma parameters see Tab.\ \ref{tab:recipes}, SF$_6$/Ar cleaning (S)]. Note that all masks initially undergo cleaning in a boiling tri-acid mixture (1:1:1 HNO$_3$:H$_2$SO$_4$:HClO$_4$, \unit[5]{ml} each) and subsequent ultrasonic cleaning in acetone and isopropanol to coarsely remove deposit and residual glue/resist.  \label{fig:maskfab}}
\end{figure}

Mechanical polishing typically forms plates with down to roughly \unit[20]{$\mu$m} thickness\,\cite{Pomorski2013}. However, to ensure mechanical stability during handling and processing, we start with plates with an initial thickness of \unit[50]{$\mu$m}. The commercially-available SCD plates have been mechanically polished by Delaware Diamond Knives or Almax Easy Lab to a roughness of R$_{a}$ < $\unit[3]{nm}$ (lateral size 2x4 or \unit[3x3]{mm$^2$}). 
We use two different purity grades of CVD SCD: electronic grade SCD (Element Six, $\mathrm{[N_s]^0< \unit[5]{ppb}}$ and $\mathrm{[B]< \unit[1]{ppb}}$) and optical grade SCD ($\mathrm{[N_s]^0< \unit[1]{ppm}}$ and $\mathrm{[B]< \unit[5]{ppb}}$) according to manufacturer specifications. We now face the challenge to remove almost \unit[50]{$\mu$m} of diamond using ICP-RIE while preserving or even enhancing surface quality. We note that wet etching of diamond is not feasible. Lithographically defined masks, e.g.\ metal or resist masks typically feature thicknesses in the micrometer range and might fully erode during the required etch. We thus employ quartz cover slips as etch masks (SPI supplies, thickness \unit[75-125]{$\mu$m}). This comparably cheap, high-purity material allows for the etching of thin membranes with high surface quality\,\cite{Appel2016, Maletinsky2012}. However, a thick mask with vertical sidewalls induces a significantly inhomogeneous etch profile [see Fig.\ \ref{fig:maskfab}(a)]: In an anisotropic plasma, highly-energetic ions hit the mask's walls at grazing angles ($>$ 80$^\circ$) and undergo specular reflection while mostly retaining their energy. The reflected ions reach the diamond surface close to the base of the mask's sidewall and add to the ion flux directly impinging in this region and thus locally enhance the etch rate. The locally enhanced etch rate induces a trench\,\cite{Hoekstra1998} which, for thin membranes, cuts through the membrane and destabilizes it.   

To avoid the risk of the membrane detaching from the surrounding diamond, a step-wise reduction of the etched area using different shadow masks during the thinning process was necessary when using water jet cut masks with vertical sidewalls\,\cite{Appel2016}. Exchanging and aligning shadow masks, however, possibly contaminates the etched area, is time consuming and highly-challenging in the case of parallel processing of membranes. To avoid this delicate step, we optimize the mask's geometry via laser cutting (see Fig.\ \ref{fig:maskfab}(b), manufacturer: Photonik Zentrum Kaiserslautern, Germany). The trenches into the quartz masks were laser cut by ultra-fast laser machining (laser system: HYPER 25, Coherent Kaiserslautern GmbH). The pulse duration was in the \unit[10]{ps} range using a laser wavelength of \unit[532]{nm}. A galvanometer scanner with focusing objective (HurrySCANII 14, SCANLAB AG) was used to deflect the laser beam with a spot diameter of \unit[12]{$\mu$m} onto the surface in order to cut the trenches and the outline of the masks precisely. By varying the cutting speed, the pulse repetition rate of the laser pulses and the processing sequences, the quality of the trenches was optimized and cracking was avoided. Fig.\ \ref{fig:maskfab}(b) depicts a typical layout of our improved mask design. We choose the outer dimensions of the mask to fully cover the SCD plate. While laser cutting allows us to almost freely choose the sidewall angle of our mask (here: $\approx \,$30$^\circ$), laser ablation creates deposit of silica debris all over the mask. Preliminary tests clearly revealed that this deposit mostly consists of amorphous silica and leads, independent of plasma chemistry, to micro-masking and roughening of the thinned diamond. Using a protective layer (sandwiching between sacrificial glass cover slips or coating with a resist layer) avoids deposit in remote areas but not in direct proximity to the laser, i.e.\ especially on the side wall itself. Removing this deposit via wet etching (KOH or HF) significantly roughened the side walls. Alternatively, we use ICP-RIE etching (Oxford, PlasmaLab100 and Sentech, ICP 500): Fig.\ \ref{fig:maskfab}(c) and (d) show scanning electron microscope (SEM) images of the mask's sidewall before and after cleaning in SF$_6$/Ar plasma. Table \ref{tab:recipes} summarizes the etch parameters. The masks have been thinned by \unit[3]{$\mu$m} [low magnification image Fig.\ \ref{fig:maskfab}(d)] or \unit[10]{$\mu$m} [inset Fig.\ \ref{fig:maskfab}(d)] in the etching. The sidewall morphology changes, indicating full removal of the deposit induced by laser cutting for both procedures. No roughening of mask or sidewalls occurs during our plasma cleaning. 
\begin{table}
\caption{\label{tab:recipes} Plasma recipes used in this work. Ar/SF$_6$ cleaning removes deposit from laser cutting from the quartz masks. The recipes used in the Sentech reactor in Basel, marked with (S), and the Oxford reactor in Saarbr\"ucken, marked with (O), slightly differ but lead to comparable results. We note that the Cl$_2$-based recipe is run on a ceramic based carrier system, while the SF$_6$-based deep etch is run on standard silicon carrier wafers.  Anode temperature is set to \unit[20]{$^{\circ}$C}. For the SF$_6$-based deep etch, the etch rate in parenthesis gives the rate at which the Ar/SF$_6$/O$_2$ erode the quartz mask.  }
\begin{tabular}{p{3cm}cccccc}
\hline
\textbf{Plasma}  & \textbf{ICP power} & \textbf{RF power} & \textbf{DC bias}& \textbf{Gas Flux} & \textbf{Pressure}&  \textbf{Etch rate}  \\
  & [W] & [W] & [V] & [sccm] & [Pa] &  [nm/min] \\
\hline
Ar/SF$_6$ & 700  & 100& 170 & SF$_6$ 10  & 1.2  & 97  \\
mask clean (O)&&&& Ar 20  &&\\
\hline
Ar/SF$_6$& 700  & 220& 150& SF$_6$ 25  & 2  & 161  \\
mask clean (S)&&&& Ar 50  &&\\
\hline
Cl$_2$-based  & 400  & 100&200 & Ar 25  & 1 &   35  \\
diamond deep etch&&&& Cl$_2$ 40  &&\\
(S) & 700  & 50 &110 & O$_2$ 60  & 1.3 &   126  \\
\hline
SF$_6$-based & 700  & 100& 170 & SF$_6$ 10  & 1.2  & 67  \\
diamond deep etch &&&& Ar 20  &&\\
(O)& 700  & 100 & 160 & Ar 15  & 1.6  & 87(51)  \\
&&&& SF$_6$ 7  &&\\
&&&& O$_2$ 22  &&\\
\hline
\end{tabular}
\end{table}
 
\section{Etching of SCD membranes \label{seclabel:memb_fab}}
Prior to etching, we subject the SCD plates to cleaning in a boiling tri-acid mixture (1:1:1 HNO$_3$:H$_2$SO$_4$:HClO$_4$, \unit[5]{ml} each)  as well as solvent cleaning (acetone, isopropanol) to remove any contamination arising from polishing. Failure to start etching on a clean surface induces micro-masking (e.g.\ needle formation) and consequently destroys the surface integrity. The SCD plates which we place on a carrier chip (silicon for SF$_6$-based processes, Al$_2$O$_3$ for Cl$_2$-based processes) are then covered with the shadow mask. We attach the mask using an adhesive (crystal bond 509) or vacuum grease (Dow Corning, high vacuum grease) to the carrier chip. We use two recipes that employ an alternating sequence of plasma types: in the SF$_6$-based approach [Oxford, Plasmalab 100, Saarbr\"ucken (SB)], we start with \unit[10]{min} Ar/SF$_6$ followed by \unit[20]{min} Ar/SF$_6$/O$_2$ plasma and cycle this sequence. In the Cl$_2$-based recipe (Sentech, SI 500 ICP-RIE etch chamber, Basel), we start with \unit[5]{min} Ar/Cl$_2$. Then, the cycling part of the recipe starts consisting of \unit[5]{min} Ar/Cl$_2$ and \unit[10]{min} O$_2$ plasma. Cycling continues until the membrane reaches the desired thickness. We terminate etching with an O$_2$-containing step. We note that to keep the temperature in the etching chamber low, we insert a cooling step of \unit[4]{mins} (Basel) or \unit[5]{mins} (SB) after each \unit[5]{mins} of etching.  To enable efficient cooling, we purge the reactor using Ar during this period (Basel: \unit[100]{sccm}, \unit[13.2]{Pa}, SB: \unit[30]{sccm}, \unit[1]{Pa}).  
In these cycling recipes, O$_2$-containing plasma steps etch SCD fast, while not inducing too fast mask erosion (see Tab.\ \ref{tab:recipes}). The O$_2$-free steps help to ensure a smooth surface. For the SF$_6$-based recipe,  SF$_6$ hinders the re-deposition of sputtered mask material in the  O$_2$-containing steps\,\cite{Tran2010}. We thus ensure a clean etching process that ensures smooth surfaces and low contamination (see below).

The starting phase of the etching procedure is particularly critical: First, mechanically polished SCD plates often contain sub-surface, structural defects due to polishing within the first micrometers below the surface\,\cite{Volpe2009}. O$_2$-containing plasmas have been found to transform such polishing damage into etch pits and thus roughen the surface\,\cite{Friel2009} which motivates starting both recipes with the O$_2$-free plasma step. Second, despite careful cleaning, transferring the plate to the etch chamber might result in minor contamination. Also from this point of view starting with the O$_2$-free plasma is advantageous as this potentially more efficiently removes contaminants either by Ar physical etching or by chemical etching. Note that the approach employed here leads to an enhanced surface quality compared to previous approaches using only Ar/SF$_6$/O$_2$ with the same ratio of gases in a more simple RIE machine\,\cite{Jung2016}.  

Our Sentech SI 500 ICP-RIE etch chamber permits \textit{in situ} etch rate measurements via a laser interferometer (SenTech SLI 670). For the processes performed in the Oxford Plasmalab 100, we perform \textit{ex situ} measurements using cross-sectional scanning electron microscopy SEM (Hitachi S800) or tip-based measurements (Veeco Dektak 150, Stylus \unit[5]{$\mu$m} radius, resolution \unit[0.067]{$\mu$m/sample} ). Etch rates of the respective plasma steps are given in Tab.\ \ref{tab:recipes}. Taking into account the necessary cooling steps, removing $\approx \,$\unit[50]{$\mu$m} of diamond leads to an overall etching process time of typically less than \unit[20]{hours} which is comparable for both approaches. 
\section{Membrane roughness and trenching \label{seclabel:memb_char}}
\begin{figure}[h!]
	\centering
	\includegraphics[width=0.9\linewidth]{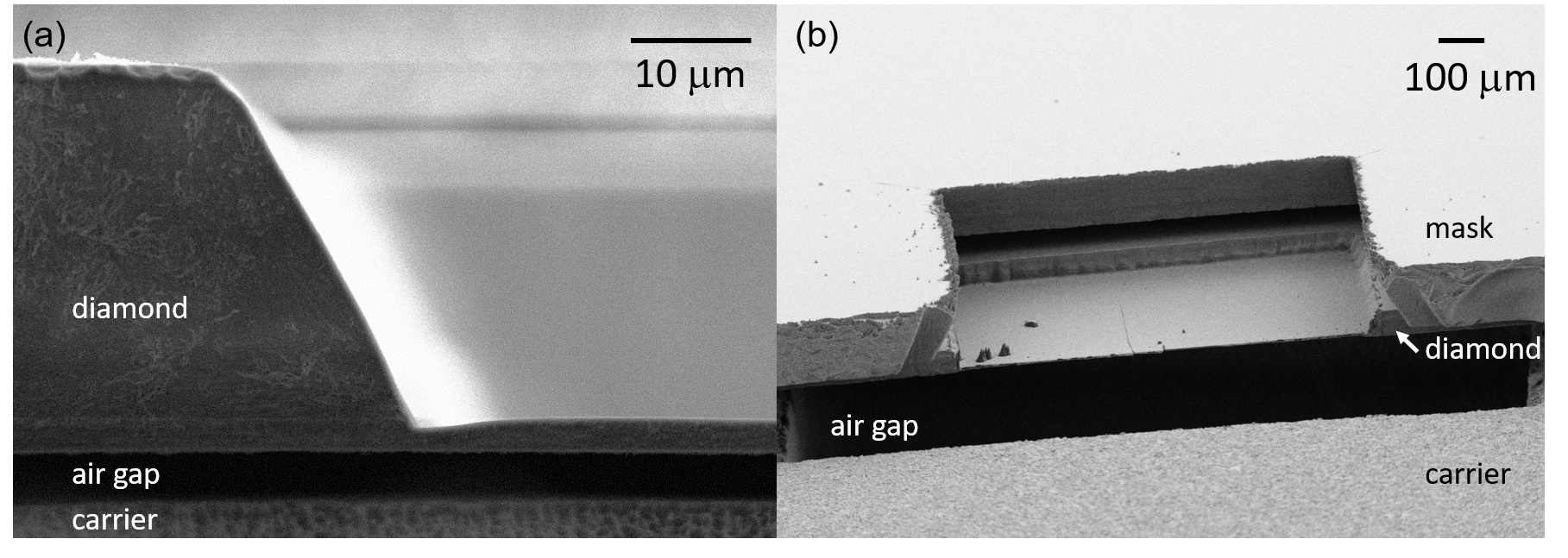}
	\caption{Electron microscopy characterization of the etched membranes: (a) Cross-sectional view of a membrane etched using our SF$_6$-based recipe. The membrane's thickness at its edge as deduced from this image is \unit[2.8]{$\mu$m}. A trench with a depth of \unit[0.8]{$\mu$m} is visible in the image. (b) View of a membrane etched using the Cl$_2$-based process (taken under $\approx\,$\unit[80]{$^\circ$}). For the membrane etched in the Cl$_2$-based process, the mask is still attached. In case of the Cl$_2$-based process, a needle wall had to be removed mechanically from the outer edge of the membrane. This process occasionally leads to cracks in the membrane. In contrast the outer edge of the membrane etched in the  SF$_6$-based process is smooth without further treatment.   \label{fig:membchar}}
\end{figure}
\begin{figure}[h!]
	\centering
	\includegraphics[width=0.9\linewidth]{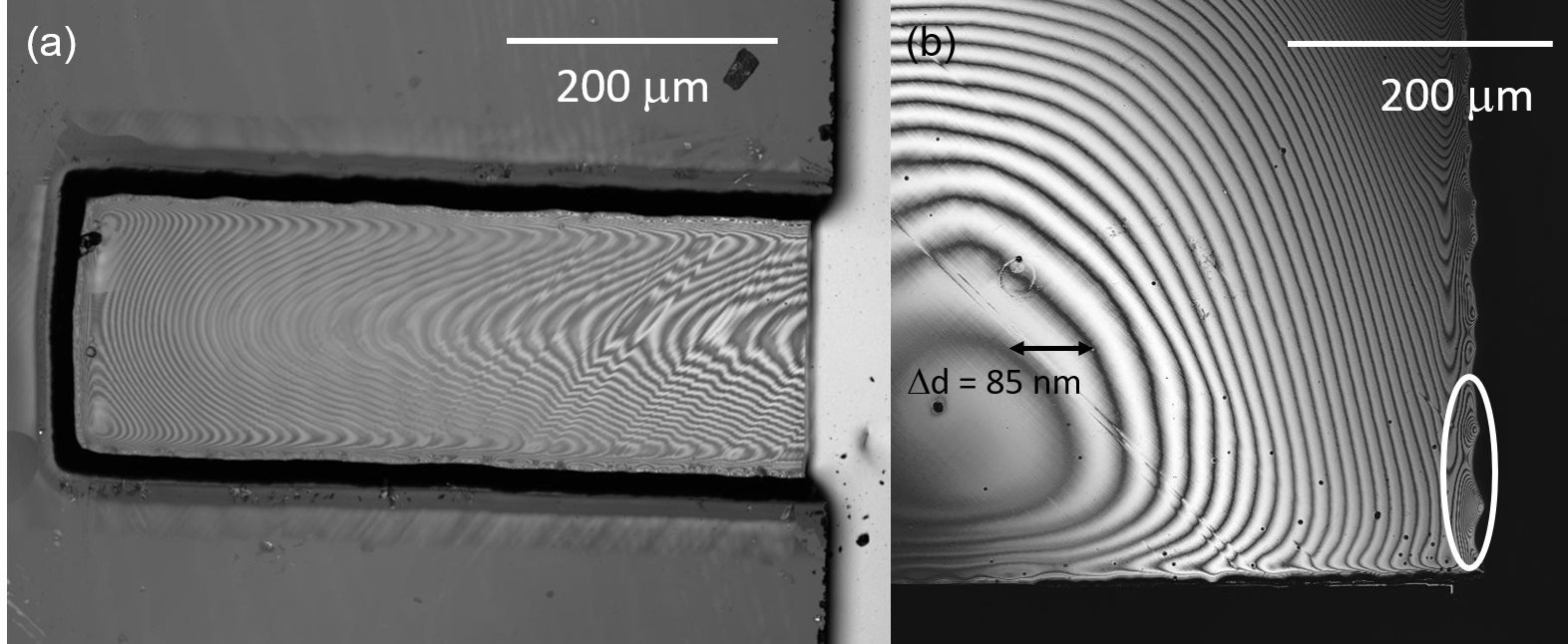}
	\caption{Interference fringes of the etched membranes recorded using a confocal laser scanning microscope (LSM): (a) Thickness homogeneity of a membrane etched with the SF$_6$-based recipe. The thickness of the membrane at the outer edge (thinnest point) is \unit[2.5]{$\mu$m} as measured using SEM. The diamond membrane shows a thickness variation of \unit[3.6]{$\mu$m} along its long edge and a variation of \unit[0.8]{$\mu$m} along its short edge direction, as.  Measured using a LEXT OLS4100 LSM (Olympus) equipped with a \unit[405]{nm} laser. (b) Membrane etched using the Cl$_2$-based process, compare Fig.\  \ref{fig:membchar}(b). The thickness of this membrane at its outer edge is \unit[2]{$\mu$m} as measured using SEM. The encircled area illustrates the local variation of the trenching as described in the text. Measured using a VK-X210 LSM (Keyence) equipped with a \unit[408]{nm} laser.  \label{fig:membchar2}}
\end{figure}

We first compare the trenching for different etch recipes, as well as different mask geometries. The trenching at the edge of the membrane is measured via cross-sectional SEM images [see Fig.\ \ref{fig:membchar}(a)] and checked via laser scanning confocal microscopy for membranes with a thickness of less than \unit[6.5]{$\mu$m}.  For both etch recipes, the optimized mask geometry as depicted in Fig.\ \ref{fig:maskfab}(b) clearly reduces the trenching. For the SF$_6$-based recipe, though showing a strong scatter of the observed trench depths, we reduce the trenching depth to less than \unit[2]{$\mu$m}. For more than half of the etched membranes, the trench depth reduces to less than \unit[1]{$\mu$m} or might even be negligible. For the Cl$_2$-based recipe, the reduction of the trenching is even more pronounced: often only a very shallow (< \unit[200]{nm}) trench forms, which is not recognizable in the SEM images. We point out, however, that the trench depth shows strong local variations: We use a confocal laser scanning microscope (LSM) to investigate these local variations of the trenches. The local brightness variation (fringes) that we record in Fig.\,\ref{fig:membchar2} corresponds to areas of destructive and constructive interference of the laser light reflected by the front and back surfaces of our membrane that arise when the focused laser is being scanned over the sample. The changes thus precisely reflect local thickness variations of our SCD membrane. From the laser wavelengths and the refractive index (n = 2.4) of SCD, we calculate that between two dark fringes the membrane thickness changes by $\approx \,$\unit[85]{nm}. For the position marked in Fig.\ \ref{fig:membchar2}(b), the trench depth locally increases to \unit[400-800]{nm} (corresponding to 5-10 fringes). Thus, the optimized mask geometry allows to fabricate stable membranes with roughly down to \unit[1]{$\mu$m} thickness.   

One technical advantage of the SF$_6$-based approach is that the outer edge of the membrane remains smooth during etching. For the Cl$_2$-based recipe, a wall of needles forms often times possibly due to re-deposition of Al$_2$O$_3$ or amorphous carbon on the sidewall of the diamond plate. The latter is reported to potentially have high etch resistance in multiple plasma chemistries \cite{McKenzie2011a}. Indeed, measurements using energy dispersive X-ray spectroscopy (EDX) revealed the presence of aluminum, oxygen, silicon and chlorine under the shadow mask close to the etched area. Removing these deposits can be straightforwardly performed using HF and is thus not influencing subsequent processing steps. However, to gain full access to the membrane, e.g.\ for resist coating, it is often necessary to remove these needles mechanically, which can break the membrane. For the SF$_6$-based approach, EDX measurements show weaker contamination with aluminum, oxygen and fluorine under the shadow mask close to the etched area. In the SF$_6$-based approach, the contaminants did not result in needle formation on the membrane's outer edge.   

We use constructive and destructive interference fringes recorded in a confocal laser scanning microscope (LSM) to investigate the thickness homogeneity of our membranes as discussed above. Figure \ref{fig:membchar2} displays two examples of thin membranes etched with the two recipes. In general, we find that the geometry of the etched membrane influences its thickness homogeneity. Especially for square membranes, the central part shows a thickness variation of less than \unit[1]{$\mu$m} over $\approx \,$\unit[200x200]{$\mu$m$^2$}. The homogeneity reached for these membranes is suitable for the fabrication of e.g.\ scanning probe nanostructures. Our new mask geometry does not seem to influence the homogeneity significantly compared to previous results\,\cite{Appel2016}.

\begin{figure}[h!]
	\centering
	\includegraphics[width=0.8\linewidth]{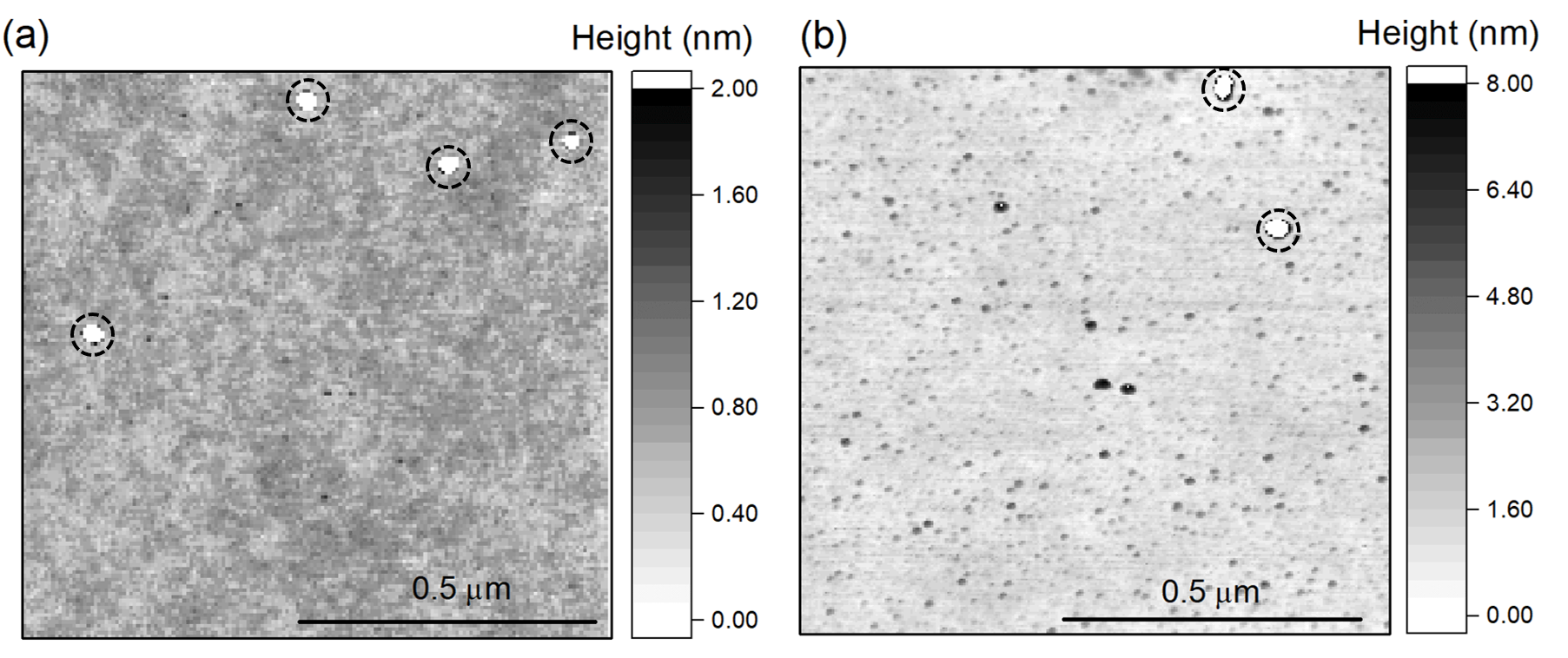}
	\caption{ (a) Surface roughness of a membrane etched with the SF$_6$-based plasma process to thickness of \unit[2.5]{$\mu$m}. Image recorded using a FastScan-ScanAsyst AFM (Bruker) in tapping mode. The four encircled bumps exceed the chosen scale bar and are \unit[9.3]{nm} high. (b) Membrane etched using the Cl$_2$-based process to a thickness of \unit[9]{$\mu$m}. Image recorded using a Dimension 3100 AFM (Bruker) in tapping mode. The two encircled bumps marked exceed the chosen scale bar and are \unit[18]{nm} high.    \label{fig:membchar3}}
\end{figure}

To quantitatively analyze the surface quality of the membranes, we use AFM measurements in tapping mode. Before etching, we find an RMS surface roughness of \unit[0.8]{nm} in accordance with manufacturer specifications. For the etched membranes, we measure a roughness of \unit[0.3]{nm} [SF$_6$-based process, area \unit[1]{$\mu$m$^2$}, optical grade, see Fig.\ \ref{fig:membchar3}(a)] and \unit[0.5]{nm} [Cl$_2$-based process, area \unit[1]{$\mu$m$^2$}, electronic grade, see Fig.\ \ref{fig:membchar3}(b)].  Though these numbers can be only estimates due to the limited resolution of the AFM and a roughness almost on the atomic scale, they clearly prove the very high surface quality of the membranes which renders them usable for many applications including photonics and nanomechanics. We note that in both processes the membranes have been smoothed during the etching process and the polishing direction that was often visible before the etch is not discernible anymore in the AFM images. 

\section{Conclusion and outlook}
We introduce a cycling etching recipe for thin SCD membrane etching. The recipe uses argon, oxygen and SF$_6$ and reaches similar performance as the previously published recipe using argon, chlorine and oxygen concerning surface quality, homogeneity and etch rate. Furthermore, the SF$_6$-based process creates membranes without needles close to the outer edge that can hinder subsequent processing.  As we avoid the use of toxic and highly-corrosive feed gases, the recipe can be operated on standard ICP reactors and is closer to processes from semiconductor technology. 
We present a new mask geometry for a shadow mask manufactured out of quartz cover slides. With these masks, we reduce the trenching of the membranes for SF$_6$-based and Cl$_2$-based processes. These new approaches potentially foster the up-scaling of SCD thin membrane formation via parallel processing of several membrane windows. 

Beyond this study, further optimization of our process will consider the charge state stability and spin coherence of color centers created close to the etched surfaces (distance $<$ \unit[10]{nm}) of our membranes. Our Cl$_2$-based recipe has already been used for sensing applications with NV centers and enables stable NV centers with roughly \unit[75]{$\mu$s} coherence time at a depth of \unit[9]{nm}\cite{Appel2016}. However, both plasma recipes introduced here use partly physical etching with a high bias voltage and argon admixture in the plasma. Such approaches allow us to reliably remove several tens of micrometers of SCD with a high etch rate while conserving smooth surfaces. Nevertheless, our highly-biased plasma recipes potentially damage the surface \cite{Oliveira2015}. Future optimization will include finishing the etching using the promising approach of a low-bias, low-damage plasma step with low etch rate \cite{Oliveira2015}. Furthermore, the surface termination of the SCD membranes is crucial for shallow color centers as it influences their charge state. We aim at controlling the surface termination, e.g.\ via dedicated plasma treatments \cite{Osterkamp2013}.
\vspace{6pt}

\acknowledgments{We gratefully acknowledge financial support through the NCCR QSIT, a competence center funded by the Swiss NSF, through the Swiss Nanoscience Institute, and through SNF Grants No.\ 142697 and 155845. EN and coworkers acknowledge funding via a NanoMatFutur grant of the German Ministry of Education and Research (FKZ13N13547) as well as a PostDoc Fellowship by the Daimler and Benz Foundation. We thank J\"org Schmauch and Daniel Mathys for performing EDX analysis, Dr.\,Rene Hensel and Susanne Selzer (INM, Saarbr\"ucken) for providing the plasma etching tool and assistance, Thomas Veit for enabling to use the AFM device, Günther Marchand for assistance with the Dektak measurements, Christoph Pauly for the introduction to the Laser Scanning Microscopes and Thomas Herrrmann (Photonikzentrum Kaiserslautern) for optimizing the laser cutting process.}


\begin{thebibliography}{-------}
\providecommand{\natexlab}[1]{#1}

\bibitem[Zaitsev(2001)]{Zaitsev2001}
Zaitsev, A.
\newblock {\em {Optical Properties of Diamond: A Data Handbook}}; Springer,
  2001.

\bibitem[Aharonovich and Neu(2014)]{Aharonovich2014a}
Aharonovich, I.; Neu, E.
\newblock Diamond Nanophotonics.
\newblock {\em Adv. Opt. Mater.} {\bf 2014}, {\em 2},~911--928.

\bibitem[Bhaskar \em{et~al.}(2017)Bhaskar, Sukachev, Sipahigil, Evans, Burek,
  Nguyen, Rogers, Siyushev, Metsch, Park, Jelezko, Lon\ifmmode~\check{c}\else
  \v{c}\fi{}ar, and Lukin]{Bhaskar2017}
Bhaskar, M.K.; Sukachev, D.D.; Sipahigil, A.; Evans, R.E.; Burek, M.J.; Nguyen,
  C.T.; Rogers, L.J.; Siyushev, P.; Metsch, M.H.; Park, H.; Jelezko, F.;
  Lon\ifmmode~\check{c}\else \v{c}\fi{}ar, M.; Lukin, M.D.
\newblock Quantum Nonlinear Optics with a Germanium-Vacancy Color Center in a
  Nanoscale Diamond Waveguide.
\newblock {\em Phys. Rev. Lett.} {\bf 2017}, {\em 118},~223603.

\bibitem[Pomorski \em{et~al.}(2013)Pomorski, Caylar, and
  Bergonzo]{Pomorski2013}
Pomorski, M.; Caylar, B.; Bergonzo, P.
\newblock Super-thin single crystal diamond membrane radiation detectors.
\newblock {\em Applied physics letters} {\bf 2013}, {\em 103},~112106.

\bibitem[Rondin \em{et~al.}(2014)Rondin, Tetienne, Hingant, Roch, Maletinsky,
  and Jacques]{Rondin2014}
Rondin, L.; Tetienne, J.P.; Hingant, T.; Roch, J.F.; Maletinsky, P.; Jacques,
  V.
\newblock Magnetometry with nitrogen-vacancy defects in diamond.
\newblock {\em Rep. Prog. Phys.} {\bf 2014}, {\em 77},~056503.

\bibitem[Bernardi \em{et~al.}(2017)Bernardi, Nelz, Sonusen, and
  Neu]{Bernardi2017}
Bernardi, E.; Nelz, R.; Sonusen, S.; Neu, E.
\newblock Nanoscale Sensing Using Point Defects in Single-Crystal Diamond:
  Recent Progress on Nitrogen Vacancy Center-Based Sensors.
\newblock {\em Crystals} {\bf 2017}, {\em 7},~124.

\bibitem[Dolde \em{et~al.}(2011)Dolde, Fedder, Doherty, N{\"o}bauer, Rempp,
  Balasubramanian, Wolf, Reinhard, Hollenberg, Jelezko, and
  Wrachtrup]{Dolde2011}
Dolde, F.; Fedder, H.; Doherty, M.W.; N{\"o}bauer, T.; Rempp, F.;
  Balasubramanian, G.; Wolf, T.; Reinhard, F.; Hollenberg, L.; Jelezko, F.;
  Wrachtrup, J.
\newblock Electric-field sensing using single diamond spins.
\newblock {\em Nat. Phys.} {\bf 2011}, {\em 7},~459--463.

\bibitem[Neumann \em{et~al.}(2013)Neumann, Jakobi, Dolde, Burk, Reuter,
  Waldherr, Honert, Wolf, Brunner, Shim, Suter, Sumiya, Isoya, and
  Wrachtrup]{Neumann2013}
Neumann, P.; Jakobi, I.; Dolde, F.; Burk, C.; Reuter, R.; Waldherr, G.; Honert,
  J.; Wolf, T.; Brunner, A.; Shim, J.H.; Suter, D.; Sumiya, H.; Isoya, J.;
  Wrachtrup, J.
\newblock High-Precision Nanoscale Temperature Sensing Using Single Defects in
  Diamond.
\newblock {\em Nano Lett.} {\bf 2013}, {\em 13},~2738--2742.

\bibitem[Teissier \em{et~al.}(2014)Teissier, Barfuss, Appel, Neu, and
  Maletinsky]{Teissier2014}
Teissier, J.; Barfuss, A.; Appel, P.; Neu, E.; Maletinsky, P.
\newblock Strain Coupling of a Nitrogen-Vacancy Center Spin to a Diamond
  Mechanical Oscillator.
\newblock {\em Phys. Rev. Lett.} {\bf 2014}, {\em 113},~020503.

\bibitem[Ali~Momenzadeh \em{et~al.}(2016)Ali~Momenzadeh, de~Oliveira, Neumann,
  Bhaktavatsala~Rao, Denisenko, Amjadi, Chu, Yang, Manson, Doherty, and
  Wrachtrup]{Momenzadeh2016}
Ali~Momenzadeh, S.; de~Oliveira, F.F.; Neumann, P.; Bhaktavatsala~Rao, D.D.;
  Denisenko, A.; Amjadi, M.; Chu, Z.; Yang, S.; Manson, N.B.; Doherty, M.W.;
  Wrachtrup, J.
\newblock Thin Circular Diamond Membrane with Embedded Nitrogen-Vacancy Centers
  for Hybrid Spin-Mechanical Quantum Systems.
\newblock {\em Phys. Rev. Applied} {\bf 2016}, {\em 6},~024026.

\bibitem[Delfaure \em{et~al.}(2016)Delfaure, Pomorski, De~Sanoit, Bergonzo, and
  Saada]{Delfaure2016}
Delfaure, C.; Pomorski, M.; De~Sanoit, J.; Bergonzo, P.; Saada, S.
\newblock Single crystal CVD diamond membranes for betavoltaic cells.
\newblock {\em Applied Physics Letters} {\bf 2016}, {\em 108},~252105.

\bibitem[Riedrich-M{\"o}ller \em{et~al.}(2015)Riedrich-M{\"o}ller, Pezzagna,
  Meijer, Pauly, M{\"u}cklich, Markham, Edmonds, and
  Becher]{RiedrichMoeller2015}
Riedrich-M{\"o}ller, J.; Pezzagna, S.; Meijer, J.; Pauly, C.; M{\"u}cklich, F.;
  Markham, M.; Edmonds, A.M.; Becher, C.
\newblock Nanoimplantation and Purcell enhancement of single nitrogen-vacancy
  centers in photonic crystal cavities in diamond.
\newblock {\em Appl. Phys. Lett.} {\bf 2015}, {\em 106},~221103.

\bibitem[Riedel \em{et~al.}(2017)Riedel, S{\"o}llner, Shields, Starosielec,
  Appel, Neu, Maletinsky, and Warburton]{Riedel2017}
Riedel, D.; S{\"o}llner, I.; Shields, B.J.; Starosielec, S.; Appel, P.; Neu,
  E.; Maletinsky, P.; Warburton, R.J.
\newblock Deterministic Enhancement of Coherent Photon Generation from a
  Nitrogen-Vacancy Center in Ultrapure Diamond.
\newblock {\em Phys. Rev. X} {\bf 2017}, {\em 7},~031040.

\bibitem[Appel \em{et~al.}(2016)Appel, Neu, Ganzhorn, Barfuss, Batzer, Gratz,
  Tsch{\"o}pe, and Maletinsky]{Appel2016}
Appel, P.; Neu, E.; Ganzhorn, M.; Barfuss, A.; Batzer, M.; Gratz, M.;
  Tsch{\"o}pe, A.; Maletinsky, P.
\newblock Fabrication of all diamond scanning probes for nanoscale
  magnetometry.
\newblock {\em Review of Scientific Instruments} {\bf 2016}, {\em 87},~063703.

\bibitem[Kleinlein \em{et~al.}(2016)Kleinlein, Borzenko, M{\"u}nzhuber, Brehm,
  Kiessling, and Molenkamp]{Kleinlein2016}
Kleinlein, J.; Borzenko, T.; M{\"u}nzhuber, F.; Brehm, J.; Kiessling, T.;
  Molenkamp, L.
\newblock NV-center diamond cantilevers: Extending the range of available
  fabrication methods.
\newblock {\em Microelectron. Eng.} {\bf 2016}, {\em 159},~70--74.

\bibitem[Maletinsky \em{et~al.}(2012)Maletinsky, Hong, Grinolds, Hausmann,
  Lukin, Walsworth, Loncar, and Yacoby]{Maletinsky2012}
Maletinsky, P.; Hong, S.; Grinolds, M.; Hausmann, B.; Lukin, M.; Walsworth, R.;
  Loncar, M.; Yacoby, A.
\newblock A robust scanning diamond sensor for nanoscale imaging with single
  nitrogen-vacancy centres.
\newblock {\em Nat. Nanotechnol.} {\bf 2012}, {\em 7},~320--324.

\bibitem[Thiel \em{et~al.}(2016)Thiel, Rohner, Ganzhorn, Appel, Neu, Kleiner,
  Koelle, and Maletinsky]{Thiel2016}
Thiel, L.; Rohner, D.; Ganzhorn, M.; Appel, P.; Neu, E.; Kleiner, R.; Koelle,
  D.; Maletinsky, P.
\newblock Quantitative nanoscale vortex-imaging using a cryogenic quantum
  magnetometer.
\newblock {\em Nat. Nanotechnol.} {\bf 2016}, {\em 11},~677.

\bibitem[Riedrich-M{\"o}ller \em{et~al.}(2012)Riedrich-M{\"o}ller, Kipfstuhl,
  Hepp, Neu, Pauly, M{\"u}cklich, Baur, M~Wandt, Wolff, Fischer, Gsell,
  Schreck, and Becher]{Riedrichmoeller2011}
Riedrich-M{\"o}ller, J.; Kipfstuhl, L.; Hepp, C.; Neu, E.; Pauly, C.;
  M{\"u}cklich, F.; Baur, A.; M~Wandt, M.; Wolff, S.; Fischer, M.; Gsell, S.;
  Schreck, M.; Becher, C.
\newblock One- and two-dimensional photonic crystal microcavities in single
  crystal diamond.
\newblock {\em Nature Nanotech.} {\bf 2012}, {\em 7},~69.

\bibitem[Piracha \em{et~al.}({2016})Piracha, Ganesan, Lau, Stacey, McGuinness,
  Tomljenovic-Hanic, and Prawer]{Piracha2016a}
Piracha, A.H.; Ganesan, K.; Lau, D.W.M.; Stacey, A.; McGuinness, L.P.;
  Tomljenovic-Hanic, S.; Prawer, S.
\newblock {Scalable fabrication of high-quality, ultra-thin single crystal
  diamond membrane windows}.
\newblock {\em NANOSCALE} {\bf {2016}}, {\em {8}},~6860--6865.

\bibitem[Lee \em{et~al.}(2013)Lee, Magyar, Bracher, Aharonovich, and
  Hu]{Lee2013}
Lee, J.; Magyar, A.; Bracher, D.; Aharonovich, I.b.; Hu, E.
\newblock Fabrication of thin diamond membranes for photonic applications.
\newblock {\em Diamond Relat. Mater.} {\bf 2013}, {\em 33},~45--48.
\newblock cited By (since 1996) 0.

\bibitem[Babinec \em{et~al.}(2011)Babinec, Choy, Smith, Khan, and
  Lon{\v{c}}ar]{Babinec2011}
Babinec, T.M.; Choy, J.T.; Smith, K.J.; Khan, M.; Lon{\v{c}}ar, M.
\newblock Design and focused ion beam fabrication of single crystal diamond
  nanobeam cavities.
\newblock {\em Journal of Vacuum Science \& Technology B, Nanotechnology and
  Microelectronics: Materials, Processing, Measurement, and Phenomena} {\bf
  2011}, {\em 29},~010601.

\bibitem[Wan \em{et~al.}(2018)Wan, Mouradian, and Englund]{Wan2018}
Wan, N.H.; Mouradian, S.; Englund, D.
\newblock Two-Dimensional Photonic Crystal Slab Nanocavities on Bulk
  Single-Crystal Diamond.
\newblock {\em arXiv preprint arXiv:1801.01151} {\bf 2018}.

\bibitem[Tao \em{et~al.}(2014)Tao, Boss, Moores, and Degen]{Tao2014}
Tao, Y.; Boss, J.; Moores, B.; Degen, C.
\newblock Single-crystal diamond nanomechanical resonators with quality factors
  exceeding one million.
\newblock {\em Nat. Commun.} {\bf 2014}, {\em 5},~3638.

\bibitem[Lee \em{et~al.}(2008)Lee, Gu, Dawson, Friel, and Scarsbrook]{Lee2008}
Lee, C.; Gu, E.; Dawson, M.; Friel, I.; Scarsbrook, G.
\newblock Etching and micro-optics fabrication in diamond using chlorine-based
  inductively-coupled plasma.
\newblock {\em Diamond Relat. Mater.} {\bf 2008}, {\em 17},~1292--1296.

\bibitem[Hoekstra \em{et~al.}(1998)Hoekstra, Kushner, Sukharev, and
  Schoenborn]{Hoekstra1998}
Hoekstra, R.J.; Kushner, M.J.; Sukharev, V.; Schoenborn, P.
\newblock Microtrenching resulting from specular reflection during chlorine
  etching of silicon.
\newblock {\em Journal of Vacuum Science \& Technology B: Microelectronics and
  Nanometer Structures} {\bf 1998}, {\em 16},~2102--2104.

\bibitem[Tran \em{et~al.}(2010)Tran, Fansler, Grotjohn, Reinhard, and
  Asmussen]{Tran2010}
Tran, D.; Fansler, C.; Grotjohn, T.; Reinhard, D.; Asmussen, J.
\newblock Investigation of mask selectivities and diamond etching using
  microwave plasma-assisted etching.
\newblock {\em Diamond and Related Materials} {\bf 2010}, {\em 19},~778--782.

\bibitem[Volpe \em{et~al.}(2009)Volpe, Muret, Omnes, Achard, Silva, Brinza, and
  Gicquel]{Volpe2009}
Volpe, P.N.; Muret, P.; Omnes, F.; Achard, J.; Silva, F.; Brinza, O.; Gicquel,
  A.
\newblock Defect analysis and excitons diffusion in undoped homoepitaxial
  diamond films after polishing and oxygen plasma etching.
\newblock {\em Diamond Relat. Mater.} {\bf 2009}, {\em 18},~1205 -- 1210.

\bibitem[Friel \em{et~al.}(2009)Friel, Clewes, Dhillon, Perkins, Twitchen, and
  Scarsbrook]{Friel2009}
Friel, I.; Clewes, S.; Dhillon, H.; Perkins, N.; Twitchen, D.; Scarsbrook, G.
\newblock Control of surface and bulk crystalline quality in single crystal
  diamond grown by chemical vapour deposition.
\newblock {\em Diam. Relat. Mater.} {\bf 2009}, {\em 18},~808--815.

\bibitem[Jung \em{et~al.}(2016)Jung, Kreiner, Pauly, M{\"u}cklich, Edmonds,
  Markham, and Becher]{Jung2016}
Jung, T.; Kreiner, L.; Pauly, C.; M{\"u}cklich, F.; Edmonds, A.M.; Markham, M.;
  Becher, C.
\newblock Reproducible fabrication and characterization of diamond membranes
  for photonic crystal cavities.
\newblock {\em physica status solidi (a)} {\bf 2016}, {\em 213},~3254--3264.

\bibitem[McKenzie \em{et~al.}(2011)McKenzie, Pethica, and Cross]{McKenzie2011a}
McKenzie, W.; Pethica, J.; Cross, G.
\newblock A direct-write, resistless hard mask for rapid nanoscale patterning
  of diamond.
\newblock {\em Diamond and Related Materials} {\bf 2011}, {\em 20},~707--710.

\bibitem[de~Oliveira \em{et~al.}({2015})de~Oliveira, Momenzadeh, Wang, Konuma,
  Markham, Edmonds, Denisenko, and Wrachtrup]{Oliveira2015}
de~Oliveira, F.F.; Momenzadeh, S.A.; Wang, Y.; Konuma, M.; Markham, M.;
  Edmonds, A.M.; Denisenko, A.; Wrachtrup, J.
\newblock {Effect of low-damage inductively coupled plasma on shallow
  nitrogen-vacancy centers in diamond}.
\newblock {\em {APPLIED PHYSICS LETTERS}} {\bf {2015}}, {\em {107}},~073107.

\bibitem[Osterkamp \em{et~al.}(2013)Osterkamp, Scharpf, Pezzagna, Meijer,
  Diemant, J\''urgen~Behm, Naydenov, and Jelezko]{Osterkamp2013}
Osterkamp, C.; Scharpf, J.; Pezzagna, S.; Meijer, J.; Diemant, T.;
  J\''urgen~Behm, R.; Naydenov, B.; Jelezko, F.
\newblock Increasing the creation yield of shallow single defects in diamond by
  surface plasma treatment.
\newblock {\em Appl. Phys. Lett.} {\bf 2013}, {\em 103},~193118.

\end{thebibliography}
\end{document}